\documentstyle[12pt]{article}
\begin{document}
\begin{center}
{\bf Self-organized criticality in self-directing walks}\\
\vspace*{1cm}
{\bf V.B.Priezzhev}\\
\vspace*{0.5cm}
Bogolubov Laboratory of Theoretical Physics, \\
Joint Institute for Nuclear Research,\\
Dubna, Moscow region, 141980 Russia\\
\vspace*{3cm}
{\bf Abstract}\\
\vspace*{0.5cm}
\end{center}
A new model of self-organized criticality is proposed. An algebra
 of operators is introduced which is similar to that used for the Abelian
 sandpile model. The structure of the configurational space is
determined and the number of recurrent states is found.

\vspace*{2cm}
Since  the introduction of self-organized criticality (SOC)
\cite{Bak}, there has been considerable interest in the study of
 cellular automata which demonstrate how power-law correlations
emerge during the evolution of extended dissipative systems.  Of
them, the Abelian sandpile model \cite{Dhar} expresses most
clearly the idea of dissipative dynamics, where a small
disturbance exceeding a threshold grows and propagates through
the system as an avalanche.

Avalanches seem to be crucial for SOC, while the significance of
a threshold is not completely clear. Indeed, some of the
characteristics of sandpiles are purely diffusive even though
one might expect more complicated behaviour due to the presence
of thresholds of stability. Thresholds in the forest fire model
\cite{Drossel} are hidden in two parameters which are the
probabilities of growth and ignitions. In the Bak-Sneppen model
\cite{Sneppen}, a threshold value appears as a result of
infinitely long evolution.

In this Letter, I propose a cellular automaton model exhibiting
SOC and containing no threshold parameters. However, the model
has a common algebraic structure with the Abelian sandpile and,
therefore, a similar structure of recurrent configurations
corresponding to the critical state. The parallel consideration
of these models should elucidate the critical dynamics of both.

Consider a two dimensional square lattice $\cal L$ of size $L
\times L$.  Each site $i$ of $\cal L$ is characterized by the
 radius-vector ${\bf r}_{i}$ with integer Cartesian coordinates
 $(x_{i},y_{i})$ and by one of the unit vectors ${\bf e}_{i}(1),{\bf
 e}_{i}(2),{\bf e}_{i}(3),{\bf e}_{i}(4)$ directed up, right,
 down or left from $i$ .  At each discrete moment of time, one
 drops a particle on a site of $\cal L$ chosen at random and
 allows it to walk by the following rules:

\begin{quotation}
(i)at each step, the particle coming to a site $j \in \cal L$ turns
 the vector ${\bf e}_{j}(\nu)$ clockwise by the right
 angle: $${\bf e}_{j}(\nu) \rightarrow {\bf e}_{j}(\mu) $$
where $\mu = \nu + 1 (mod 4) $.

(ii) performs the unit step in the direction ${\bf e}_{j}(\mu)$
 to the neighbour site $j^{'}$:
$$ {\bf r}_{j^{'}} = {\bf r}_{j} + {\bf e}_{j}(\mu)$$

(iii) if the walk reaches a boundary site $j$  and
 the new position $j^{'}$ is outside  the lattice,
 the particle leaves the system.
\end{quotation}

The walk is assumed to be quick enough to be completed by the
 next discrete moment of time.  As a result of the walk, a
configuration of vectors $C$ specified by a unit vector ${\bf
 e}_{i}(\nu)$ on each lattice site $i \in \cal L$ transforms
 into a new configuration $C^{'}$ which can  generally differ from
 $C$ by directions of vectors on sites visited by the particle.

To describe the transformation resulting from dropping
 a particle on the site $i$ , we define the operator $a_{i}$
acting on $C$ and producing $C^{'}$:
\begin{equation}
\label{1}
a_{i} C = C^{'}
\end{equation}

{\bf Theorem 1}\hspace{0.5cm} The operator $a_{i}$ exists, that
 is, for any $C$ the walk started at $i$  never enters a
 non-trivial cycle.

Proof: If a cycle contains a boundary site, the walk visits this
 site more than four times. Since one of these visits has as a
 consequence a step  outside  the lattice, no cycle can
 contain a boundary site. Next, consider a site one step away
 from the boundary. It cannot be visited in a cycle as,in this case,
 a particle must hit  one of the the boundary sites. By
 induction, there can be no cycle containing an arbitrary site
 of the lattice and, therefore, the operator $a_{i}$ exists.

The operators $a_{i}$ all commute. This property enables one to
 construct an Abelian group quite similar to that defined by
 Dhar \cite{Dhar} for the sandpile cellular automata.

{\bf Theorem 2}\hspace{0.5cm} For arbitrary sites $i$ and $j$
 and for any configuration of vectors $C$
\begin{equation}
\label{2}
a_{i}a_{j} C = a_{j}a_{i} C
\end{equation}

Proof: Consider the updating procedure $a_{i} C = C^{'}$ as a
 sequence of elementary steps
 $\alpha_{j_{n}}...\alpha_{j_{2}}\alpha_{j_{1}}$ with $j_{1} =
 i$. The operator $\alpha_{j}$ corresponds to the rotation of
 the vector $ {\bf e}_{j}(\nu) \rightarrow {\bf e}_{j}(\mu)
 $,$\mu = \nu + 1 (mod 4) $ and a consequent single step in the
 direction ${\bf e}_{j}(\mu) $. Then $a_{i}a_{j} C $ can be
 written in the form $$(\prod \alpha^{(2)}_{j})(\prod
\alpha^{(1)}_{k}) C$$ where subscripts (1) and (2) refer to
different particles. If $j \not = j^{'}$, the operators
$\alpha^{(1)}_{j}$ and $\alpha^{(2)}_{j^{'}}$ commute.  If $j
= j^{'}$, they also commute due to identity of particles.
Therefore, $a_{i}$ and $a_{j}$ commute.

The commutativity rule enables us to consider several walks
simultaneously updating them concurrently.

As usual,in the theory of Markov chains, we divide the set of all
configurations  $\{C\}$ into two subsets, recurrent and
transient. The first subset denoted by $\{R\}$ includes those
configurations which can be obtained from an arbitrary
configuration by a sequential action by operators $a_{i}$. It
follows from the definition that the subset $\{R\}$ is closed
under multiple action by operators $a_{i}$. Once the system gets
into $\{R\}$, it never gets out under subsequent evolution.  All
nonrecurrent configurations are called transient and form the
subset $\{T\}$ which is the complement of the set $\{ R \}$.

Similarly to the avalanche operators of the sandpile model, the
 operator $a_{i}$ has a unique inverse.The proof of this
 statement can be carried out in a way very close to the
 sandpile construction \cite{Creutz}.

{\bf Theorem 3}\hspace{0.5cm} For any recurrent configuration $C
 \in \{ R \}$, there exists a unique $$ (a_{i}^{-1} C) \in \{ R
\}$$ such that
\begin{equation}
\label{3}
a_{i}( a_{i}^{-1} ) C =  C
\end{equation}
Proof: Consider a standard recurrent configuration $C^{*}$ which
can be chosen as the set of parallel vectors ${\bf e}_{i}(1)$ on
all $i \in \cal L$.

Construct an identity operator $E C^{*} = C^{*}$. To this end,
we take the  product of boundary operators $a_{i}$ ($i \in B$)
multiplied by the product of corner operators $a_{i}$ ($i \in
A$)
\begin{equation}
\label{4}
\prod \limits_{i\in B}\prod \limits_{i\in A} a_{i}
\end{equation}
where $B$ is the set of all boundary sites and $A$ is the set of
corner sites. Evidently, the operator (\ref{4}) does not change
the standard configuration since it is nothing but a sequence of
four successive rotations of arrows in all rows and columns of
the lattice. Besides (\ref{4}), we can find another
representation of $E$ , having a nonzero degree of $a_{i}$ at
any $i \in \cal L$.

For instance, let us construct an identity operator $E$ having
operators $a_{i}$ in the next neighbours of the boundary sites.
Replace $a_{i}$ by  $\alpha _{i}$ and consider the product

\begin{equation}
\label{5}
\left(\prod \limits_{i\in B}\prod \limits_{i\in A} \alpha_{i}\right)^{4}
\end{equation}
Let $n_{i}$,$i \in \cal L$ be the occupation numbers of
particles resulting from $C^{*}$ after the action by the operator
(\ref{5}).Each site a step away from the boundary receives at
least one particle. Thus, the operator
\begin{equation}
\label{6}
\prod \limits_{i} a_{i}^{n_{i}}
\end{equation}
is the identity operator for $C^{*}$ having $n_{i} > 0$ at each
next neighbour of edges. Repeating this procedure, we can
construct the identity operator $E(n_{i})$ having $n_{i} > 0$ at
an arbitrary chosen site $i \in \cal L$.

Now, drop a particle on the configuration $C^{*}$ to obtain a
new recursive configuration $a_{i} C^{*}$. The operator
\begin{equation}
\label{7}
P_{1} = E(n_{i}-1)
\end{equation}
being combined with $a_{i}$ gives the identity operator again:
\begin{equation}
\label{8}
P_{1} a_{i} C^{*} = C^{*}
\end{equation}
By definition, any recurrent configuration $C$ can be
obtained from $C^{*}$:
\begin{equation}
\label{9}
C = P_{2} C^{*}
\end{equation}
where $P_{2}$ is a product of the $a_{i}$. Using the
commutativity property, we have
\begin{equation}
\label{10}
C = P_{2}P_{1}a_{i} C^{*} = a_{i} P_{2}P_{1} C^{*}
\end{equation}
The intermediate result $P_{2}P_{1} C^{*}$ is the seeking
configuration $(a_{i}^{-1} C)$.

The proof of uniqueness doesn't differ from that for sandpiles
\cite{Creutz}. Repeating the construction (\ref{10}), we can
find a sequence $C_{n} \in \{ R \}$ such that
\begin{equation}
\label{11}
(a_{i})^{n} C_{n} = C
\end{equation}
Since the total number of configurations doesn't exceed
$4^{L^{2}}$, this sequence must enter a loop of length $m > 1$.
This loop must contain $C$ as $C \in \{ R \}$ is attainable from
an arbitrary point of the loop. We have $a_{i}^{m} C = C$. Then
$a_{i}^{m-1} C = a_{i}^{-1}C$ is the unique inverse.

As all recurrent configurations can be obtained from an
arbitrary one by the successive acting by operators $a_{i}$, we
can represent any $C \in \{ R \}$ in the form
\begin{equation}
\label{12}
C = \prod \limits_{i\in \cal L}(a_{i})^{n_{i}} C^{*}
\end{equation}
The $L^{2}$-dimensional vector $\bf n$ labels the recurrent
configurations. We can note that the operator $a_{i}^{4}$
returns the vector ${\bf e}_{i}(\nu)$ to the former position and
initiates a motion of four particles at neighbouring sites of
$i$. Therefore, the operator
$a_{i}^{4}a_{j_{1}}^{-1}a_{j_{2}}^{-1}a_{j_{3}}^{-1}a_{j_{4}}^{-1}$,
where $j_{1},j_{2},j_{3},j_{4}$ are the nearest neighbours of $i$,
doesn't change the initial configuration and is actually the identity
operator $E$. Using the Laplacian matrix $\Delta$ with elements

\vspace*{0.5cm}
\begin{tabular}{llll}
\hspace*{2cm}&$\Delta_{i,j} = 4 $ & $i = j$&\hspace*{3cm}\\
&$\Delta_{i,j} = -1 $ & $i,j$ nearest neighbours&\\
&$\Delta_{i,j} = 0 $ & otherwise &\\
\end{tabular}\\
\vspace*{0.5cm}
one can write down the identity operator in the form
\begin{equation}
\label{13}
E = \prod \limits_{j} a_{j}^{\Delta_{ij}}
\end{equation}
Eq. (\ref{13}) shows that two vectors $\bf n$ and $\bf n^{'}$
label the same configuration of arrows if the difference between
them is $\sum\nolimits_{j} m_{j} \Delta_{ij}$ where $m_{j}$ are
integers. The $L^{2}$-dimensional space $\{\bf n\}$ has a
periodic structure with the elementary cell of the form of a
hyper- parallelepiped with the base edges $\Delta_{ij}, j =
1,2,...,L^{2}$. Thus, the number of non-equivalent recurrent
configurations is
\begin{equation}
\label{14}
N = det \Delta
\end{equation}
which is  Kirhhoff's formula for  spanning trees
\cite{Harary} and  Dhar's formula for sandpiles \cite{Dhar}.

The similarity to trees is not accidental. By  construction,
vectors involved into rotation don't form closed cycles if
the motion is over. Due to
finiteness of the lattice, each vector takes part in the motion
some time or other. Each time the updating procedure is over,
the collection of vectors reproduces the set of bonds of a tree.
Thus, starting from an arbitrary noncorrelated set of vectors,
we come to the set of spanning trees that are characterized by
power-law correlations between different sites.

The correspondence with sandpiles is not surprising as well. The
algebra of the operators $a_{i}$ completely coincides with that
of avalanche operators of the Abelian sandpile model
\cite{Dhar}. Moreover, the identity operator (\ref{13}) has the
same form for both the models . This is the reason why the
numbers of recurrent configurations coincide.

Continuing the analogy between self-directing walks and
sandpiles one can find the expected number $G_{ij}$ of full
rotations of the vector at site $j$, due to the particle dropped
at $i$ \cite{Dhar}. During the walk, the expected number of
steps outside $j$ is $\Delta_{jj} G_{ij}$ whereas $-\sum_{k
\not=j}G_{ik}\Delta_{kj}$ is the average flux into $j$. Equating
both fluxes one gets

\begin{equation}
\label{15}
\sum\limits_{k}G_{ik} \Delta_{kj} = \delta_{ij}
\end{equation}
or
\begin{equation}
\label{16}
G_{ij} = [ \Delta^{-1} ]_{ij}
\end{equation}

The close analogy with sandpiles calls for a definition of avalanches
in our model. The first step after landing of the particle generally leads
to emegence of a cyclic configuration of arrows. As a result,
the system leaves the recurrent set . It is natural to define the avalanche
as a process of restoration of the recurrent state.It corresponds to
successive rotations from the beginning of the motion up to the moment
when an acyclic configuration is restored for the first time and the
structure of the spanning tree is reconstructed. The number of steps $n$ is
the duration of an avalanche, the number of different
visited sites $s$ is its
size. If the first step doesn't lead to a cyclic configuration, we put
$ n = 1, s = 1 $.

When a closed loop appears on the given lattice, a branch of the dual
tree gets disconnected on the dual lattice. The probability distribution
$P(s)$ of disconnected clusters follows the power law \cite{MDM}
\begin{equation}
\label{17}
P(s) \sim \frac{1}{s^{11/8}}
\end{equation}
where $s$ is the number of lattice sites belonging to a cluster.
Since the number of steps $n$ which are neccessary to restore the tree is
proportional to $s$, one can expect the similar power law for the
avalanche distribution $P(n)$.

The proposed model admits a natural generalization on an arbitrary graph
and arbitrary order of numeration of unit vectors directed to nearest
neighbours of a given site. The main result (\ref{14}) remains unchanged
where $\Delta$ should be defined as the Laplacian matrix of the given
graph.

\end{document}